\documentclass[]{IEEEtran}

\usepackage{epsfig}
\usepackage{subfigure}
\usepackage{verbatim}
\usepackage{amsfonts}
\usepackage{amssymb}
\usepackage{amsmath}
\usepackage{cuted}
\usepackage{enumerate}
\usepackage{graphicx}
\usepackage{multirow}

\usepackage{color}
\usepackage{soul}

\begin{document}


\title{An Energy Consumption Model \\for IEEE 802.11ah WLANs}


\author{\IEEEauthorblockN{Albert Bel~\IEEEmembership{Member,~IEEE}, Toni Adame, and Boris Bellalta~\IEEEmembership{Senior Member,~IEEE}\\}
\IEEEauthorblockA{Department of Information and Communication Technologies \\ Universitat Pompeu Fabra, Barcelona}
\thanks{Manuscript received December 1, 2012; revised September 17, 2014. Corresponding author: A. Bel (email: albert.bel@upf.edu)}}

\markboth{Journal of \LaTeX\ Class Files,~Vol.~13, No.~9, September~2014}%
{Shell \MakeLowercase{\textit{et al.}}: Bare Demo of IEEEtran.cls for Journals}

\IEEEtitleabstractindextext{%
\begin{abstract}
One of the main challenges when designing a new self-powered wireless sensor network (WSN) technology is the vast operational dependence on its scarce energy resources. Therefore, a thorough identification and characterisation of the main energy consumption processes may lay the foundation for developing further mechanisms aimed to make a more efficient use of devices' batteries. This paper provides an energy consumption model for IEEE 802.11ah WLANs operating in power saving mode, which are expected to become one of the technology drivers in the development of the Internet of Things (IoT) in the next years. Given the network characteristics, the presented analytical model is able to provide an estimation of the average energy consumed by a station as well as to predict its battery lifetime. Once the model has been validated, we use it to obtain the optimal IEEE 802.11ah power saving parameters in several IoT key scenarios, validating that the parameters provided by the IEEE 802.11ah Task Group are already a very good choice.



\end{abstract}

\begin{IEEEkeywords}
IEEE 802.11ah, WLANs, M2M, WSNs, Power Saving Mechanisms.
\end{IEEEkeywords}}

\maketitle

\IEEEdisplaynontitleabstractindextext

%
\IEEEpeerreviewmaketitle

\section{Introduction}
\label{introduction}

The growing use of Machine to Machine (M2M) communications \cite{machina2014the} envisages a future where personal and business decision making will be increasingly based on the information provided by these unattended systems. Their autonomous, scattered, ubiquitous and non-invasive nature facilitates the procedure of obtaining environmental data large amounts of sensors, but at the same time supposes a technological challenge, as most of their conforming devices are strongly conditioned by processing, memory and, particularly, energy constraints. 

Indeed, neither of the two current IoT/M2M players (cellular networks and WSNs) has yet been able to produce a prevailing technology with such considerations, thus fostering the appearance of new low capability communication standards such as IEEE 802.11ah \cite{bellalta2015next}. Conceived as an amendment of the consolidated and well-known IEEE 802.11 Wireless Local Area Network (WLAN) technology, this under-development amendment will offer a competitive long-range solution in the sub 1 GHz band for very large WSNs (i.e., $>$ 8K devices) with low power consumption and short-burst data transmission requirements ($<$ 100 Bytes) \cite{aust2012ieee}. To achieve that, besides modifying the IEEE 802.11ac PHY layer to operate in the sub 1 GHz band, IEEE 802.11ah includes new power management mechanisms \cite{khorov2015survey}. One of them, called \textit{TIM and page segmentation} \cite{IEEE802.11-12/1084r4}, extends the IEEE 802.11 power saving mechanism \cite{IEEE802.11-2012} and distributes network stations and channel resources according to a novel hierarchical method. Basically, energy consumption of a STA is reduced by limiting the number of possible contenders in its corresponding TDMA-like transmission period. 

The analytical characterization of energy consumption in IEEE 802.11ah WLANs has been already considered by the research community. In \cite{zheng2013performance}, an analytical model to characterize the performance of the \textit{TIM and page segmentation} scheme is proposed, although no calculations of energy consumption are included. On the contrary, \cite{liu2013power} surveys the performance (collision probability, delay, and battery lifetime) of IEEE 802.11ah networks with periodic traffic while \cite{raeesi2014performance} predicts their saturation throughput and energy efficiency assuming known collision and error probabilities. As for IEEE 802.11ah simulations, a model to calculate the maximum number of stations using power saving mechanisms is presented in \cite{adame2013}, a performance assessment of its power saving mechanism is included in \cite{adame2014IEEE}, and a novel low-consuming channel access mechanism is proposed in \cite{EW_subm}.

This paper presents an analytical model for the energy consumption in an IEEE 802.11ah WLAN, where all elements of the \textit{TIM and page segmentation} scheme (including signalling beacons, number of stations per group, transmission periods, and so forth) are taken into consideration to compute it. In addition, the model accuracy has been evaluated by comparing the model predictions with the results presented in \cite{adame2014IEEE}, where the energy consumption of an IEEE 802.11ah WLAN in four typical M2M scenarios (agriculture monitoring, smart metering, industrial automation and animal monitoring) is evaluated by simulation. The obtained results reflect the similarity between the proposed model and the simulations, thus proving its effectiveness when predicting the energy consumption and the average lifetime of an IEEE 802.11ah WLAN. Moreover, once the model is validated, it has been used to optimize several IEEE 802.11ah parameters in terms of energy consumption and probability of successfully transmit a packet.

The remainder of this paper is organised as follows: Section \ref{s_model} provide the main parameters and assumptions of the considered IEEE 802.11ah network while Section \ref{e_model} details the equations of our analytical model. Its performance is evaluated in Section \ref{evaluation} and the optimization of model variables is provided in Section \ref{optimization}. Lastly, Section \ref{conclusions} presents the conclusions and discusses open challenges.


\section{System Model} \label{s_model}

\subsection{IEEE 802.11ah WLAN Operation}

IEEE 802.11ah extends the IEEE 802.11 power saving mechanism (PSM) \cite{IEEE802.11-2012} by using a scheme called \textit{TIM and page segmentation}, which reduces the time a STA is competing for the channel and increases its sleeping periods. 

IEEE 802.11ah introduces a novel hierarchical method to build groups of stations depending on an association identifier. This hierarchical distribution of stations into groups, called TIM groups, is used not only for organizational purposes but also for scheduling signalling and allocating available channel resources, allowing stations to enter in an sleep mode during non-traffic periods. Hence, a STA only wakes up at predefined moments to listen to the beacons, which are the following:

\begin{enumerate}
    \item DTIM (Delivery Traffic Indication Map) beacons. They must be listened to by all STAs and inform about which TIM groups have pending data in the AP and also about multicast and broadcast messages.
    \item TIM (Traffic Indication Map) beacons. Between two DTIM beacons, there are as many TIM beacons as TIM groups. Each TIM beacon informs a group of STAs about which specific ones have pending data in the AP. 
\end{enumerate}

After listening to DTIM beacons, transmitted every $T$ seconds, a STA with pending data from the AP or pending data to transmit, will wake up to listen the corresponding TIM beacon, from where the STAs obtain information of the downlink and uplink RAW segments. Hence, a STA with packets to transmit or receive is only awake in its TIM period, remaining in sleeping mode otherwise.

\begin{figure*}[t!!]
\centering
\includegraphics[width=6.9in]{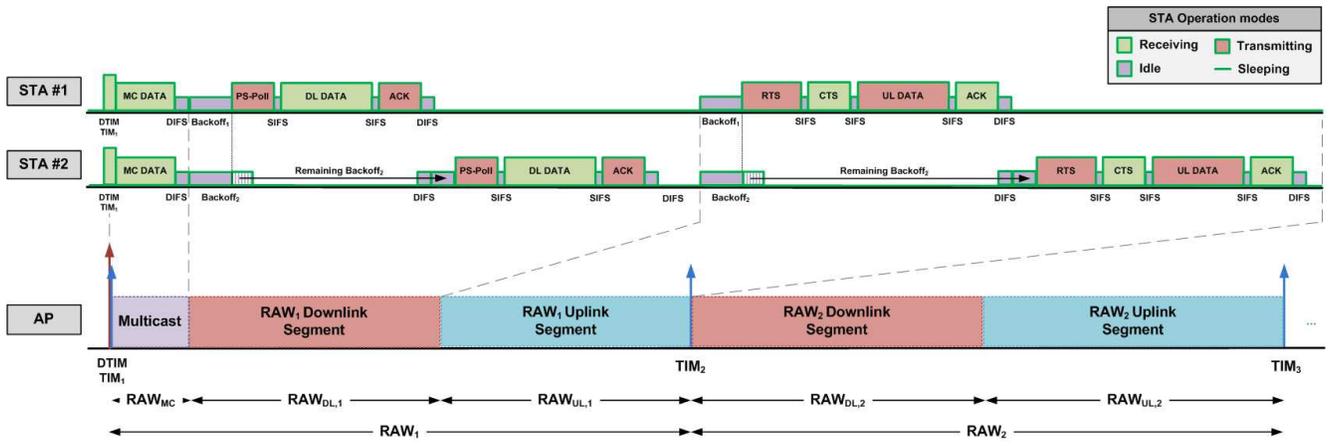}
\caption{Beacon signalling of the AP and transmission procedures of TIM STAs in IEEE 802.11ah WLANs}
\label{fig:segments}
\end{figure*}

In addition, the time between consecutive TIMs contains a restricted access window (RAW) formed by one downlink (DL) segment, one uplink (UL) segment, and one multicast (MC) segment placed immediately after each DTIM beacon. Distribution of beacons and RAW slots is shown in Figure \ref{fig:segments}. As we only consider the existence of TIM stations, the time between TIM beacons is only distributed between the RAW downlink and uplink periods. 

In the present paper, where only STAs using the \textit{TIM and page segmentation} scheme have been considered, the channel access combines an AP-centralized time period allocation system with the distributed coordination function (DCF) medium-access technique within those periods. The data transmission procedures for both the downlink and uplink cases are shown in Figure \ref{fig:segments} and detailed as follows:

\begin{enumerate}
    \item \textit{Downlink}: When a STA has a data packet pending to receive, it will be informed first by the inclusion of its TIM group in the DTIM bitmap and later by its own inclusion in the TIM bitmap. To initiate the reception of its packet, the STA will send a PS-Poll frame in its assigned RAW downlink segment. 
    \item \textit{Uplink}: Whenever a station wants to send an uplink message to the AP, it must first wait for its corresponding RAW uplink segment. Once within it, the STA will start the data transmission by using basic access or RTS/CTS mechanism.
\end{enumerate}

\subsection{Scenario} \label{scenario}

\begin{figure}[t!!]
\centering
\includegraphics[width=3.0in]{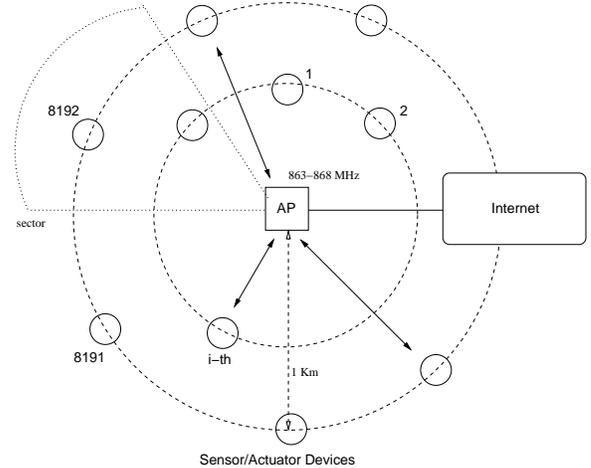}
\caption{IEEE 802.11ah scenario considered in this paper.}
\label{fig:Wifi_M2M}
\end{figure}

As shown in Figure \ref{fig:Wifi_M2M}, an IEEE 802.11ah WLAN that consists of $N_{\text{STA}}$ STAs randomly distributed over a given area and a single AP placed at its center is considered. By applying STA sectorization, all nodes are able to detect transmissions from any other node in their TIM group, and therefore, collisions with hidden nodes are not considered. The number of sectors is the same as the number of TIM groups. In \cite{Park12} and \cite{Galcey12}, the authors discuss about the hidden node problem in IEEE 802.11ah networks, and they provide possible solutions to avoid this problem, e.g., the sectorization of the STAs through the TIM groups or the use of information from an AP to spread out uplink transmissions over a period of time, thus eliminating the effects of hidden nodes.

\subsubsection{Channel model}

STAs and the AP communicate at rate $R_{dB}(d)$, which depends on the distance between both devices and the environment's path loss. To compute the value of $R_{dB}(d)$ we follow the approach presented in \cite{Hazmi2012} for both indoor and outdoor scenarios:\footnote{For further details, we refer the reader to \cite{Hazmi2012}, where the calculation of $R_{dB}(d)$ is explained in detail.}
\small
\begin{equation}
R_{dB}(d) = P_{\text{TX}}+G_{\text{TX}}-\text{PL}(d)-\text{FM}(d)+G_{\text{RX}}- \left( \frac{E_b}{N_0} \right)_{dB}-N_0 
\label{data_rate}
\end{equation}
\normalsize
where $P_{TX}$ is the transmission power, $G_{TX}$ and $G_{RX}$ are the antenna gains at the transmitter and receiver, respectively, PL$(d)$ is the path loss, FM$(d)$ is the fade margin, and $\left( \frac{E_b}{N_0} \right)_{dB}$ value depends on the modulation and coding rate used. The different existing modulations and coding rates, as well as the rest of parameters used in this paper to compute the path loss, are shown in Table \ref{tab:phy_model}.

Transmitted packets can suffer from transmission errors with probability $p_e$. The value of $p_e$ is assumed to be constant regardless the distance between the STA and the AP. This assumption is justified by the use of multiple transmission rates. We consider that the modulation and coding rate are adapted to compensate for the change in signal-to-noise ratio with the goal of keeping $p_e$ constant. Moreover, it is only applied to DATA packets in both downlink and uplink. Other packets are therefore considered error-free.

 \begin{table}
 \begin{center}
 
 \resizebox {9cm }{!}{
  \begin{tabular}{|c|l|c|c|c|c|}
  \hline
  \textbf{Parameter} & \textbf{Description} & \multicolumn{4}{|c|}{\textbf{Value}} \\
  \hline
  f & Carrier frequency & \multicolumn{4}{|c|}{900 MHz}  \\
  \hline
  $d_{\text{BP}}$ & Breakpoint distance & \multicolumn{4}{|c|}{5 m} \\
  \hline
  $P_{\text{TX}}$ & Transmission power & \multicolumn{4}{|c|}{0 dBm} \\
  \hline
 $G_{\text{TX}}$ & Transmission gain & \multicolumn{4}{|c|}{0 dBi}  \\
 \hline
 $G_{\text{RX}}$ & Reception gain & \multicolumn{4}{|c|}{3 dBi} \\
 \hline
 $T_0$ & Receiver temperature & \multicolumn{4}{|c|}{293 K} \\
 \hline
 $N_0$ & Noise figure & \multicolumn{4}{|c|}{3 dB}\\
 \hline
 $B$ & Bandwidth & \multicolumn{4}{|c|}{1 MHz}\\
 \hline
 $L$ & Data packet size & \multicolumn{4}{|c|}{100 bytes}\\
 \hline
 PER & Packet error rate & \multicolumn{4}{|c|}{10\%} \\
 \hline
 \multirow{2}{*}{FM} & \multirow{2}{*}{Fade margin} & \multicolumn{4}{|c|}{Outdoor   12.82 dB}\\
 
 & & \multicolumn{4}{|c|}{Indoor    3.84 dB}\\
 \hline
 \multirow{11}{*}{Mod} & \multirow{11}{*}{Modulation} & mode &modulation & code rate & data rate (R)\\
 \cline{3-6}
 & & MCS 0  & BPSK & 1/2 & 300 kbps\\
\cline{3-6}
 & & \multirow{2}{*}{MCS 1-2} & \multirow{2}{*}{QPSK} & 1/2 & 600 kbps\\
 
 & & & & 3/4 & 900 kbps\\
\cline{3-6}
 & & \multirow{2}{*}{MCS 3-4}&\multirow{2}{*}{16-QAM} &  1/2 &1200 kbps\\
 & & & & 3/4 & 1800 kbps\\
\cline{3-5}
 
 & & \multirow{3}{*}{MCS 5-7}& \multirow{3}{*}{64-QAM} & 2/3 & 2400 kbps \\
 & & & & 3/4 & 2700 kbps\\
 & & & & 5/6 & 3000 kbps\\
\cline{3-6}
 
 & & \multirow{2}{*}{MCS 8-9}& \multirow{2}{*}{256-QAM} & 3/4 & 3600 kbps \\
 & & & & 5/6 & 4000 kbps\\
\cline{3-6}
 \hline

 \end{tabular}}
 \caption{Parameters of the channel propagation and IEEE 802.11ah PHY models} 
\label{tab:phy_model}
 \end{center}
 \end{table}

\subsubsection{Traffic model}

The probability of having a $\psi \in \{\text{DL},\text{UL}\}$ packet available for transmission in a DTIM interval ($p_{\psi}$) depends on the expected generation time between two consecutive packets ($\text{E}[T_{\text{pck}}]$). Since we focus in low traffic load scenarios, we assume in all cases that $T \ll T_{\text{pck}_{\psi}}$, allowing us to simply compute $p_{\psi}$ as follows:
\begin{eqnarray}\label{Eq:p_psi}
p_{\psi} = \min\left(1,\frac{T}{\text{E}[T_{\text{pck}_{\psi}}]} \right)\\ \nonumber
\label{p_packet}
\end{eqnarray}

In those conditions, almost all packet transmissions are completed in the same DTIM interval in which they were generated. Therefore, to keep the presented energy consumption model as simple as possible, we assume that packets unable to be transmitted in the DTIM in which they were generated are discarded. As we will discuss in next Sections, not considering the complex queueing dynamics in the AP and in every STA simplifies the complexity of model without compromising its accuracy and applicability.




\section{Energy Consumption Model} \label{e_model}

\begin{table*}[t!]
\begin{center}
\begin{tabular}{|c|l|c}
\hline
$T$ & Time between two consecutive DTIM beacons \\ 
\hline
$N_{\text{STA}}$ & Number of STAs \\ 
\hline 
$N_{P}$ & Number of pages\\ 
 & $N_{P}=\lceil{\frac{N_{\text{STA}}}{2048}}\rceil$ \\
\hline 
$N_{\text{TIM}}$ & Number of TIM groups \\ 
\hline 
$r$ & Station data rate\\ 
\hline 
$r_{\text{min}}$ & Minimum network data rate\\ 
\hline 
$\beta_{\psi}$ & Proportion of DL/UL traffic\\
\hline
$T_{\text{DTIM}}$ & DTIM Beacon time \cite{adame2013}\\ 
 & $T_{\text{DTIM}}=\frac{L_{\text{DTIM}}}{r_{\text{min}}}=\frac{25+ \left(11+\frac{17}{4} \cdot N_{\text{TIM}} + \frac{256}{N_{\text{TIM}}}\right) \cdot N_{P}}{r_{\text{min}}}$\\
\hline 
$T_{\text{TIM}}$ & TIM Beacon time \cite{adame2013} \\ 
 & $T_{\text{TIM}}=\frac{L_{\text{TIM}}}{r_{\text{min}}}=\frac{25+\left(10+\frac{256}{N_{\text{TIM}}}\right) \cdot N_{P}}{r_{\text{min}}}$\\
\hline 
$T_{\text{RAW}_{\psi}}$ & RAW time \cite{adame2013}\\
 & $T_{\text{RAW}_{\psi}} = \frac{1}{N_{\text{TIM}}} \cdot [(\frac{T}{N_{\text{TIM}} \cdot N_{P}}-T_{\text{MC}}-T_{\text{DTIM}}) \cdot \beta_{\psi}] + \frac{N_{\text{TIM}}-1}{N_{\text{TIM}}} \cdot [(\frac{T}{N_{\text{TIM}} \cdot N_{P}}-T_{\text{TIM}}) \cdot \beta_{\psi}]$\\
\hline
$N_{\text{STA}_{\psi}}$ & Number of STAs per TIM group \\ 
 & $N_{\text{STA}_{\psi}}=\frac{N_{\text{STA}}}{N_{\text{TIM}}}$\\
\hline 
$N_{\text{STA}_{\psi}^{*}}$ & Number of STAs per TIM group with pending $\psi$ traffic (i.e., contenders) \\ 
 & $N_{\text{STA}_{\psi}}^{*}=N_{\text{STA}_{\psi}} \cdot p_{\psi}$\\
\hline 
$N_{\psi}$ & Maximum number of $\psi$ packets in a DTIM period, as defined in \cite{adame2013} \\ 
\hline 
$p_{\psi}$ & Probability of a STA having a pending $\psi$ packet\\ 
\hline 
$p_{s_{i,j}}^{(\psi)}$ & Probability of a STA having a successful transmission with $i$ collisions and $j$ errors\\ 
\hline 
$p_{ns_{i,j}}^{(\psi)}$ & Probability of a STA not having a successful transmission with $i$ collisions and $j$ errors\\ 
\hline 
$p_{w_{i,j}}^{(\psi)}$ & Probability of a STA not crossing its corresponding $RAW_{\psi}$ segment boundary\\ 
\hline 
$T_{i,j}^{\text{op.mode}(\psi)}$ & Time consumed in the corresponding operation mode\\
 & for a $\psi$ transmission with $i$ collisions and $j$ errors\\ 
\hline 
$m_{\text{col}}$ & Retransmission limit for an PS\_ POLL (DL transmission)\\ 
 & or an RTS frame (UL transmission) due to collisions\\ 
\hline 
$m_{\text{err}}$ & Retransmission limit for a DATA packet due to errors\\ 
\hline 
$t_\text{{slot}}$ & Duration of an IEEE 802.11ah time slot\\ 
\hline 
$\text{CW}_{\text{min}}$ & Minimum value of contention window\\ 
\hline 
$\text{CW}_{\text{max}}$ & Maximum value of contention window\\ 
\hline 
$T_{\psi}$ & Time of a complete $\psi$ transmission\\ 
 & $T_{\text{DL}} = \frac{L_{\text{PS\_POLL}}}{r} + T_{\text{SIFS}} + \frac{L_{\text{DATA}}}{r} + T_{\text{SIFS}} + \frac{L_{\text{ACK}}}{r} + T_{\text{DIFS}}$\\
 & $T_{\text{UL}} = \frac{L_{\text{RTS}}}{r} + T_{\text{SIFS}} + \frac{L_{\text{CTS}}}{r} + T_{\text{SIFS}} + \frac{L_{\text{DATA}}}{r} + T_{\text{SIFS}} + \frac{L_{\text{ACK}}}{r} + T_{\text{DIFS}}$\\
\hline 
$T_{c_{\psi}}$ & Time spent during a collision\\ 
 & $T_{c_{\text{DL}}} = \frac{L_{\text{PS\_POLL}}}{r} + T_{\text{DIFS}}$\\
 & $T_{c_{\text{UL}}} = \frac{L_{\text{RTS}}}{r} + T_{\text{DIFS}}$\\
\hline 
$T_{e_{\psi}}$ & Time spent during an error\\ 
 & $T_{e_{\text{DL}}} = \frac{L_{\text{PS\_POLL}}}{r} + T_{\text{SIFS}} + \frac{L_{\text{DATA}}}{r} + T_{\text{DIFS}}$  \\
 & $T_{e_{\text{UL}}} = \frac{L_{\text{RTS}}}{r} + T_{\text{SIFS}} + \frac{L_{\text{CTS}}}{r} + T_{\text{SIFS}} + \frac{L_{\text{DATA}}}{r} + T_{\text{DIFS}}$  \\
\hline 
\end{tabular}
\caption{Main IEEE 802.11ah energy consumption model parameters} 
\label{tab:par_definition}
\end{center}
\end{table*}

The energy consumption model proposed in this paper takes as starting point the work of \cite{raeesi2014performance}, and reformulates it by including the \textit{TIM and page segmentation} elements that characterize the channel access of an IEEE 802.11ah TIM STA. 

As shown in Figure \ref{fig:segments}, within a DTIM period, a STA can perform the following actions: 
\begin{itemize}
    \item Listen to a DTIM beacon
    \item Listen to a TIM beacon    
    \item Receive a multicast (MC) packet
    \item Receive a downlink (DL) packet
    \item Transmit an uplink (UL) packet
\end{itemize}

To carry out these actions, the IEEE 802.11ah transceiver uses its different operation modes for determined time periods: receiving ($t_{\text{RX}}$), transmitting ($t_{\text{TX}}$), idle ($t_{\text{ID}}$), and sleeping ($t_{\text{SL}}$). 

Hence, the energy consumed by an IEEE 802.11ah STA (without considering data processing or sensor operation) is obtained by multiplying the time a transceiver is expected to be in each of its operation modes by the corresponding power consumption of each mode, and is given by:
\begin{equation} 
E=P_{\text{RX}} \cdot t_{\text{RX}} + P_{\text{TX}} \cdot t_{\text{TX}} + P_{\text{ID}} \cdot t_{\text{ID}} + P_{\text{SL}} \cdot t_{\text{SL}} 
\label{e_total}
\end{equation}
In the following, we will calculate the energy spent by a STA during a DTIM period according to the fraction of time it remains in each operation mode. Table \ref{tab:par_definition} may be used from now on as a reference, since it lists the main parameters considered in the model and their definition. 


\subsection{Consumption in the receiving state}

The time a STA is in the receiving state is given by \eqref{Eq:c_rx}, where all the $T_{i,j}^{\text{RX}(\psi)}$ durations are computed as follows: 
\begin{align} \label{eq:tij_rx}
    T_{i,j}^{\text{RX}(\psi)} = \alpha \cdot T_{\text{DATA}}+ \beta \cdot T_{\text{CTS}} + \gamma \cdot T_{\text{ACK}}
\end{align}
with the values of $\alpha$, $\beta$, and $\gamma$ shown in Table \ref{Tbl:tij_rx}. 

The energy consumed in receiving mode, i.e., \eqref{Eq:c_rx}, includes the following situations:

\begin{figure*}[t!]
\small
\begin{eqnarray} \label{Eq:c_rx}
t_{\text{RX}} &=& \underbrace{T_{\text{DTIM}}}_{\text{(a)}}+ \underbrace{\frac{N_{\text{TIM}}-1}{N_{\text{TIM}}} \cdot \left(p_{\text{DL}_{\text{TIM}}} \cup p_{\text{UL}}\right) \cdot T_{\text{TIM}}}_{\text{(b)}}+\underbrace{p_{\text{MC}} \cdot T_{\text{DATA}}}_{\text{(c)}}+ \\ 
&+& \underbrace{p_{\text{DL}} \cdot \left(\sum^{m_{\text{col}}-1}_{i=0}\sum^{m_{\text{err}}-1}_{j=0}p_{s_{i,j}}^{(\text{DL})} \cdot T_{i,j}^{\text{RX}(\text{DL})}+\sum^{m_{\text{col}}-1}_{i=0}p_{ns_{i,m_{\text{err}}}}^{(\text{DL})} \cdot T_{i,m_{\text{err}}}^{\text{RX}(\text{DL})} +\sum^{m_{\text{err}}-1}_{j=0}p_{ns_{m_{\text{col}},j}}^{(\text{DL})} \cdot T_{m_{\text{col}},j}^{\text{RX}(\text{DL})}+\sum^{m_{\text{col}}-1}_{i=0}\sum^{m_{\text{err}}-1}_{j=0}\left(1-p_{w_{i,j}}^{(\text{DL})}\right) \cdot T_{i,j}^{\prime \text{RX}(\text{DL})}\right)}_{\text{(d)}}+ \nonumber \\
&+& \underbrace{p_{\text{UL}} \cdot \left(\sum^{m_{\text{col}}-1}_{i=0}\sum^{m_{\text{err}}-1}_{j=0}p_{s_{i,j}}^{(\text{UL})} \cdot T_{i,j}^{\text{RX}(\text{UL})}+\sum^{m_{\text{col}}-1}_{i=0}p_{ns_{i,m_{\text{err}}}}^{(\text{UL})} \cdot T_{i,m_{\text{err}}}^{\text{RX}(\text{UL})}+\sum^{m_{\text{err}}-1}_{j=0}p_{ns_{m_{\text{col}},j}}^{(\text{UL})} \cdot T_{m_{\text{col}},j}^{\text{RX}(\text{UL})}+\sum^{m_{\text{col}}-1}_{i=0}\sum^{m_{\text{err}}-1}_{j=0}\left(1-p_{w_{i,j}}^{(\text{UL})}\right) \cdot T_{i,j}^{\prime \text{RX}(\text{UL})}\right) \nonumber}_{\text{(e)}}
\end{eqnarray}
\end{figure*}
\normalsize

\begin{table*}[t!]
\centering
\begin{tabular}{|c|c|c|c|c|c|c|c|c|}
\hline
• & \multicolumn{4}{|c|}{Downlink reception (term d in (\ref{Eq:c_rx}))} & \multicolumn{4}{|c|}{Uplink transmission (term e in (\ref{Eq:c_rx}))} \\
\hline 
• & $T_{i,j}^{\text{RX}(\text{DL})}$ & $T_{i,m_{\text{err}}}^{\text{RX}(\text{DL})}$ &  $T_{m_{\text{col}},j}^{\text{RX}(\text{DL})}$ & $T_{i,j}^{\prime \text{RX}(\text{DL})}$ & $T_{i,j}^{\text{RX}(\text{UL})}$ & $T_{i,m_{\text{err}}}^{\text{RX}(\text{UL})}$ & $T_{m_{\text{col}},j}^{\text{RX}(\text{UL})}$ & $T_{i,j}^{\prime \text{RX}(\text{UL})}$\\ 
\hline 
$\alpha$ & $j+1$ & $m_{\text{err}}$ & $j$ & $j$ & $0$ & $0$ & $0$ & $0$\\ 
\hline
$\beta$ & $0$ & $0$ & $0$ & $0$ & $j+1$ & $m_{\text{err}}$ & $j$ & $j$\\ 
\hline 
$\gamma$ & $0$ & $0$ & $0$ & $0$ & $1$ & $0$ & $0$ & $0$ \\ 
\hline 
\end{tabular}
\caption{$\alpha$, $\beta$, and $\gamma$ values for (\ref{eq:tij_rx})}\label{Tbl:tij_rx} 
\end{table*}

\begin{enumerate}[(a)]
\item \textit{DTIM Beacon transmission.} Every DTIM beacon must be listened by all TIM STAs, since they contain all necessary information to send/receive data to/from the AP.
\item \textit{TIM Beacon transmission.} A STA listens to its corresponding TIM beacon with probability $p_{\text{DL}_{\text{TIM}}} \cup p_{\text{UL}} = p_{\text{DL}_{\text{TIM}}} + p_{\text{UL}} - p_{\text{DL}_{\text{TIM}}} \cdot p_{\text{UL}}$, where $p_{\text{DL}_{\text{TIM}}}= 1 - (1-p_{\text{DL}})^{N_{\text{STA}_{\text{DL}}}}$ is the probability the AP has announced in the last DTIM beacon that it has downlink traffic addressed to the STA's corresponding TIM group and $p_{\text{UL}}$ is the probability the STA has pending data to transmit.

\item \textit{Multicast transmission.} As observed in the receiving procedure in Figure \ref{fig:segments}, there is a multicast RAW segment placed immediately after each DTIM beacon. STAs must remain in the receiving state during this segment to receive a multicast packet previously signalled in the DTIM beacon. 
\item \textit{Downlink data packet transmission.} To receive a data packet, a STA remains in the receiving state for a certain time period called $T_{i,j}^{\text{RX}(\text{DL})}$, \textcolor{black}{where $i$ is the number of collisions, $j$ is the number of errors, RX is the operation mode and DL is the traffic flow}. More specifically, this time is computed as a combination of four factors: the probability of successfully listening to the corresponding data packet $(p_{s_{i,j}})$ and the probabilities of the packet being dropped due to errors $(p_{ns_{i,m_{\text{err}}}})$, collisions with other contenders $(p_{ns_{m_{\text{col}},j}})$ or having crossed the $\text{\text{RAW}}_{\psi}$ boundary $(1-p_{w_{i,j}})$.

\item \textit{Uplink data packet transmission.} To receive the CTS and the ACK corresponding to a successful transmission, a STA remains in the receiving state for a certain time period called $T_{i,j}^{\text{RX}(\text{UL})}$. Similarly as in the previous case, this time also depends on the ability of accessing the channel, which can be affected by errors, collisions, and the $\text{RAW}_{\psi}$ size.
\end{enumerate}


\subsection{Consumption in the transmitting state}

The time a STA is in the transmitting state is given by \eqref{Eq:c_tx}, where all the $T_{i,j}^{\text{TX}(\psi)}$ durations are computed as follows: 
\begin{align} \label{eq:tij_tx}
T_{i,j}^{\text{TX}(\psi)} &=& \alpha \cdot T_{\text{PS\_ POLL}}+ \beta \cdot T_{\text{ACK}} + \gamma \cdot T_{\text{RTS}} + \delta \cdot T_{\text{DATA}}
\end{align}
with the values of $\alpha$, $\beta$, $\gamma$, and $\delta$ shown in Table \ref{Tbl:tij_tx}. 

The energy consumed in transmitting mode, i.e., \eqref{Eq:c_tx}, includes the following situations:

\begin{enumerate}[(a)]
\item \textit{Downlink reception.} To complete a successful data reception, STAs also have to send a PS-POLL and an ACK frame. The length of the time period in the transmitting state $(T_{i,j}^{\text{TX}(\text{DL})})$ varies as a function of PS-POLL sending success, which depends in turn on transmission errors, collisions and the $\text{RAW}_{\psi}$ size.

\item \textit{Uplink transmission.} Finally, it must also take into account the time a STA remains in the transmitting state due to the sending of RTS and DATA packets $(T_{i,j}^{\text{TX}(\text{UL})})$. In this case, both transmissions can be affected by errors, although only the sending of the RTS frame can suffer collisions.
\end{enumerate}

\begin{figure*}[t!]
\small
\begin{eqnarray} \label{Eq:c_tx}
t_{\text{TX}} &=& \underbrace{p_{\text{DL}} \cdot \left(\sum^{m_{\text{col}}-1}_{i=0}\sum^{m_{\text{err}}-1}_{j=0}p_{s_{i,j}}^{(\text{DL})} \cdot T_{i,j}^{\text{TX}(\text{DL})}+\sum^{m_{\text{col}}-1}_{i=0}p_{ns_{i,m_{\text{err}}}}^{(\text{DL})} \cdot T_{i,m_{\text{err}}}^{\text{TX}(\text{DL})}+\sum^{m_{\text{err}}-1}_{j=0}p_{ns_{m_{\text{col}},j}}^{(\text{DL}ç)} \cdot T_{m_{\text{col}},j}^{\text{TX}(\text{DL})}+\sum^{m_{\text{col}}-1}_{i=0}\sum^{m_{\text{err}}-1}_{j=0}\left(1-p_{w_{i,j}}^{(\text{DL})}\right) \cdot T_{i,j}^{\prime \text{TX}(\text{DL})}\right)}_{\text{(a)}}+  \nonumber \\
&+& \underbrace{p_{\text{UL}} \cdot \left(\sum^{m_{\text{col}}-1}_{i=0}\sum^{m_{\text{err}}-1}_{j=0}p_{s_{i,j}}^{(\text{UL})} \cdot T_{i,j}^{\text{TX}(\text{UL})}+\sum^{m_{\text{col}}-1}_{i=0}p_{ns_{i,m_{\text{err}}}}^{(\text{UL})} \cdot T_{i,m_{\text{err}}}^{\text{TX}(\text{UL})}+\sum^{m_{\text{err}}-1}_{j=0}p_{ns_{m_{\text{col}},j}}^{(\text{UL})} \cdot T_{m_{\text{col}},j}^{\text{TX}(\text{UL})}+\sum^{m_{\text{col}}-1}_{i=0}\sum^{m_{\text{err}}-1}_{j=0}\left(1-p_{w_{i,j}}^{(\text{DL})}\right) \cdot T_{i,j}^{\prime \text{TX}(\text{UL})}\right) }_{\text{(b)}}
\end{eqnarray}
\end{figure*}
\normalsize

\begin{table*}[t!]
\centering
\begin{tabular}{|c|c|c|c|c|c|c|c|c|}
\hline
• & \multicolumn{4}{|c|}{Downlink reception (a)} & \multicolumn{4}{|c|}{Uplink transmission (b)} \\
\hline 
• & $T_{i,j}^{\text{TX}(\text{DL})}$ & $T_{i,m_{\text{err}}}^{\text{TX}(\text{DL})}$ & $T_{m_{\text{col}},j}^{\text{TX}(\text{DL})}$ & $T_{i,j}^{\prime \text{TX}(\text{DL})}$ & $T_{i,j}^{\text{TX}(\text{UL})}$ & $T_{i,m_{\text{err}}}^{\text{TX}(\text{UL})}$ & $T_{m_{\text{col}},j}^{\text{TX}(\text{UL})}$ & $T_{i,j}^{\prime \text{TX}(\text{UL})}$\\ 
\hline 
$\alpha$ & $i+j+1$ & $i+m_{\text{err}}$ & $m_{\text{col}+j}$ & $i+j$ & $0$ & $0$  & $0$ & $0$ \\ 
\hline
$\beta$ & $1$ & $0$ & $0$ & $0$ & $0$ & $0$ & $0$ & $0$\\ 
\hline 
$\gamma$ & $0$ & $0$ & $0$ & $0$ & $i+j+1$ & $i+m_{\text{err}}$ & $m_{\text{col}+j}$ & $i+j$ \\ 
\hline 
$\delta$ & $0$ & $0$ & $0$ & $0$ & $j+1$ & $m_{\text{err}}$ & $j$ & $j$\\ 
\hline 
\end{tabular}
\caption{$\alpha$, $\beta$, $\gamma$, and $\delta$ values for (\ref{eq:tij_tx})}\label{Tbl:tij_tx} 
\end{table*}



\subsection{Consumption in the idle state}

The time a STA is in the idle state is given by \eqref{Eq:c_id}, where all the $T_{i,j}^{\text{ID}(\psi)}$ durations are computed as follows: 

\begin{align} \label{eq:tij_id}
&T_{i,j}^{\text{ID}(\psi)}  \approx  \alpha \cdot T_{\text{DIFS}}+ \beta \cdot T_{\text{SIFS}} + \\ 
& + t_{\text{slot}} \cdot \left(\sum^{\gamma}_{k=0}\frac{\min \left\lbrace 2^k \cdot \left(\text{CW}_{\text{min}}+1\right),\text{CW}_{\text{max}}+1\right\rbrace }{2}\right) + \nonumber \\
& + \left(\frac{N_{\text{STA}_{\psi}}^{*}}{2} \right) \cdot ((1-p_{c_{\psi}}) \cdot (1-p_{e_{\psi}}) \cdot T_{\psi} + \nonumber \\ 
&+p_{c_{\psi}} \cdot T_{c_{\psi}} + (1-p_{c_{\psi}}) \cdot p_{e_{\psi}} \cdot T_{e_{\psi}} ) \nonumber
\end{align}
with the values of $\alpha$, $\beta$, and $\gamma$ shown in Table \ref{Tbl:tij_id}. 

The energy consumed in idle mode, i.e., \eqref{Eq:c_id}, includes the following situations:

\begin{enumerate}[(a)]
\item \textit{Multicast reception.} After receiving a multicast packet, all STAs go to sleep except those from the first TIM group to which the AP has pending data to send, staying in the idle state for the duration of the subsequent DIFS period.
\item \textit{Downlink reception.} Time in the idle state $(T_{i,j}^{\text{ID}(\text{DL})})$ is modelled as an addition of DIFS, SIFS, backoff and waiting periods due to the data reception procedure of other TIM group contenders. In the model it is assumed that, on average, any STA transmits after the channel has been occupied by $\frac{N_{\text{STA}_{\psi}}^{*}}{2}$ contenders, which in turn can have experienced collisions or errors. The effect of the time-limited $\text{RAW}_{\psi}$ segment is also considered.
\item \textit{Uplink transmission.} Similarly, time in the idle state $(T_{i,j}^{\text{ID}(\text{UL})})$ is computed as an addition of DIFS, SIFS, backoff and waiting periods due to the data transmission procedure of other $\frac{N_{STA_{\psi}}^{*}}{2}$ contenders.
\end{enumerate}

\begin{figure*}[t!]
\small
\begin{eqnarray} \label{Eq:c_id}
t_{\text{ID}} & \approx & \underbrace{p_{\text{MC}} \cdot T_{\text{DIFS}}}_{\text{(a)}} + \\
&+& \underbrace{p_{\text{DL}} \cdot \left(\sum^{m_{\text{col}}-1}_{i=0}\sum^{m_{\text{err}}-1}_{j=0}p_{s_{i,j}}^{(\text{DL})} \cdot T_{i,j}^{\text{ID}(\text{DL})}+\sum^{m_{\text{col}}-1}_{i=0}p_{ns_{i,m_{\text{err}}}}^{(DL)} \cdot T_{i,m_{\text{err}}}^{\text{ID}(\text{DL})}+\sum^{m_{\text{err}}-1}_{j=0}p_{ns_{m_{\text{col}},j}}^{(\text{DL})} \cdot T_{m_{\text{col}},j}^{\text{ID}(\text{DL})}+\sum^{m_{\text{col}}-1}_{i=0}\sum^{m_{\text{err}}-1}_{j=0}\left(1-p_{w_{i,j}}^{(\text{DL})}\right)\cdot T_{i,j}^{\prime \text{ID}(\text{DL})}\right)}_{\text{(b)}}+  \nonumber\\
&+& \underbrace{p_{\text{UL}} \cdot \left(\sum^{m_{\text{col}}-1}_{i=0}\sum^{m_{\text{err}}-1}_{j=0}p_{s_{i,j}}^{(\text{UL})} \cdot T_{i,j}^{\text{ID}(\text{UL})}+\sum^{m_{\text{col}}-1}_{i=0}p_{ns_{i,m_{\text{err}}}}^{(\text{UL})} \cdot T_{i,m_{err}}^{\text{ID}(\text{UL})}+\sum^{m_{\text{err}}-1}_{j=0}p_{ns_{m_{\text{col}},j}}^{(\text{UL})} \cdot T_{m_{\text{col}},j}^{\text{ID}(\text{UL})}+\sum^{m_{\text{col}}-1}_{i=0}\sum^{m_{\text{err}}-1}_{j=0}\left(1-p_{w_{i,j}}^{(\text{UL})}\right)\cdot T_{i,j}^{\prime \text{ID}(\text{UL})}\right) \nonumber}_{\text{(c)}}
\end{eqnarray}
\end{figure*}
\normalsize

\begin{table*}[t!]
\centering
\begin{tabular}{|c|c|c|c|c|c|c|c|c|}
\hline 
• & \multicolumn{4}{|c|}{Downlink reception (b)} & \multicolumn{4}{|c|}{Uplink transmission (c)} \\
\hline 
• & $T_{i,j}^{\text{ID}(\text{DL})}$ & $T_{i,m_{\text{err}}}^{\text{ID}(\text{DL})}$ & $T_{m_{\text{col}},j}^{\text{ID}(\text{DL})}$ & $T_{i,j}^{\prime \text{ID}(\text{DL})}$ & $T_{i,j}^{\text{ID}(\text{UL})}$ &  $T_{i,m_{\text{err}}}^{\text{ID}(\text{UL})}$  & $T_{m_{\text{col}},j}^{\text{ID}(\text{UL})}$ & $T_{i,j}^{\prime \text{ID}(\text{UL})}$ \\ 
\hline 
$\alpha$ & $i+j+1$ & $i+m_{\text{err}}$ & $m_{\text{col}}+j$ & $i+j$ & $i+j+1$ & $i+m_{\text{err}}$ &  $m_{\text{col}}+j$ & $i+j$\\ 
\hline
$\beta$ & $j+2$ & $m_{\text{err}}$ & $j$ & $j$ & $2j+3$ & $2m_{\text{err}}$ & $2j$ & $2j$ \\ 
\hline 
$\gamma$ & $i+j$ & $i+m_{\text{err}}-1$ & $m_{\text{col}}+j-1$ & $i+j$ & $i+j$ & $i+m_{\text{err}}-1$ & $m_{\text{col}}+j-1$ & $i+j$\\ 
\hline 
\end{tabular}
\caption{$\alpha$, $\beta$, and $\gamma$ values for (\ref{eq:tij_id})}\label{Tbl:tij_id} 
\end{table*}


\subsection{Successful transmission probability}

A common element in the previous equations is the successful transmission probability of a STA after $i$ collisions and $j$ errors without crossing its corresponding $\text{RAW}_{\psi}$ segment boundary. It is noted as $p_{s_{i,j}}^{(\psi)}$ and defined by \eqref{psij}:


\begin{align} \label{psij}
&p_{s_{i,j}}^{(\psi)}=  \nonumber \\ & {i+j \choose i} \cdot 
p_{c_{\psi}}^i \cdot p_{e_{\psi}}^j \cdot \left(1-p_{c_{\psi}} \right)^{j+1}\cdot \left(1-p_{e_{\psi}}\right)\cdot p_{w_{i,j}}^{(\psi)}
\end{align}
where $p_{w_{i,j}}^{(\psi)}$ is computed as in (\ref{Eq:pw}). 

Similarly, the probability \eqref{pnsij} of a STA not having a successful transmission after $i$ collisions and $j$ errors without crossing its corresponding $\text{RAW}_{\psi}$ segment boundary is:

\begin{eqnarray}
p_{ns_{i,j}}^{(\psi)}=  {i+j \choose i} \cdot
p_{c_{\psi}}^i \cdot p_{e_{\psi}}^j \cdot \left(1-p_{c_{\psi}} \right)^{j} \cdot p_{w_{i,j}}^{(\psi)}
\label{pnsij}
\end{eqnarray}

The function which models the $\text{RAW}_{\psi}$ boundary crossing $(p_{w_{i,j}}^{(\psi)})$ compares the channel occupation time of contender stations with the size of the current $\text{RAW}_{\psi}$ segment and is defined as:


\begin{figure*}[t!]
\begin{eqnarray}\label{Eq:pw}
p_{w_{i,j}}^{(\psi)}= \left \{ 
\begin{matrix} 
1 & \mbox{if} \ \frac{N_{\text{STA}_{\psi}}^{*}}{2} \cdot \left((1-p_{c_{\psi}}) \cdot (1-p_{e_{\psi}}) \cdot T_{\psi} + p_{c_{\psi}} \cdot T_{c_{\psi}} + (1-p_{c_{\psi}}) \cdot p_{e_{\psi}} \cdot T_{e_{\psi}} \right)+ i \cdot T_{c_{\psi}} + j \cdot T_{e_{\psi}} + T_{\psi} \le T_{\text{RAW}_{\psi}}
\\ 0 & \mbox{otherwise}
\end{matrix}\right. 
\end{eqnarray}
\end{figure*}

The collision probability \eqref{pc} of a PS\_POLL frame in a DL transmission procedure or of a RTS frame in an UL transmision procedure within a determined TIM period containing $N_{\text{STA}_{\psi}}^{*}$ active STAs is given by: 

\begin{align} \label{pc}
p_{c_{\psi}} & \approx 1-\left(1-p_{\psi}\cdot\frac{1}{\text{CW}_{\text{min}}} \right)^{N_{\text{STA}_{\psi}}-1} 
\end{align}
where we have assumed that $N_{\text{STA}_{\psi}} \geqslant 1$ in the observed TIM group.



\subsection{Consumption in the sleeping state}
Finally, a STA remains asleep when not being in any other state as computed as follows:

\begin{equation} \label{c_sl}
t_{\text{SL}} = T-t_{\text{RX}}-t_{\text{TX}}-t_{\text{ID}}
\end{equation}


\section{Model Evaluation} \label{evaluation}

\begin{table}[t!]
\begin{center}
\begin{tabular}{|c|l|}
\hline
$t_{\text{simulation}}$ & $T$ = $1.6 \ s$\\ 
\hline 
$N_{\text{TIM}}$ & $8$\\ 
\hline 
$T_{\text{SIFS}}$ & $160 \ \mu s$\\ 
\hline 
$T_{\text{DIFS}}$ & $264 \ \mu s$\\ 
\hline 
$t_{\text{slot}}$ & $52 \ \mu s$\\ 
\hline
$\text{CW}_{\text{min}}$ & $16$\\ 
\hline 
$\text{CW}_{\text{max}}$ & $1024$\\ 
\hline 
$m_{\text{col}}$ & $7$\\
\hline 
$m_{\text{err}}$ & $1$\\ 
\hline 
$L_{\text{DATA}}$ & $100$ bytes\\ 
\hline 
$L_{\text{PS\_POLL}}$ & $14$ bytes\\ 
\hline 
$L_{\text{ACK}}$ & $14$ bytes\\ 
\hline 
$L_{\text{RTS}}$ & $20$ bytes\\ 
\hline 
$L_{\text{CTS}}$ & $14$ bytes\\ 
\hline 
$p_{e_{\text{DL}}}$ & $0$\\
\hline
$p_{e_{\text{UL}}}$ & $0.1$\\
\hline
\end{tabular}
\caption{Main simulation parameters for IEEE 802.11ah MAC layer} 
\label{tab:par_value}
\end{center}
\end{table}

This section provides a comparative analysis between the proposed IEEE 802.11ah energy consumption analytical model and the results obtained from simulating a fully connected IEEE 802.11ah WLAN in MATLAB. The simulator accurately reproduces the system model introduced in Section \ref{s_model}. However, differently from the model, the simulator accumulates in a buffer all those packets not transmitted in the TIM period in which they have been generated, giving the opportunity to be sent in other TIM periods. Hence, comparing the simulation results with the ones obtained from the model will allow us to quantify the impact of such assumption. Moreover, the results presented in this section will be helpful to evaluate the performance of the \textit{TIM and page segmentation mechanism} included in the IEEE 802.11ah amendment. 


\subsection{Comparison of different traffic patterns}

\begin{figure}
\centering
\includegraphics[width=9cm]{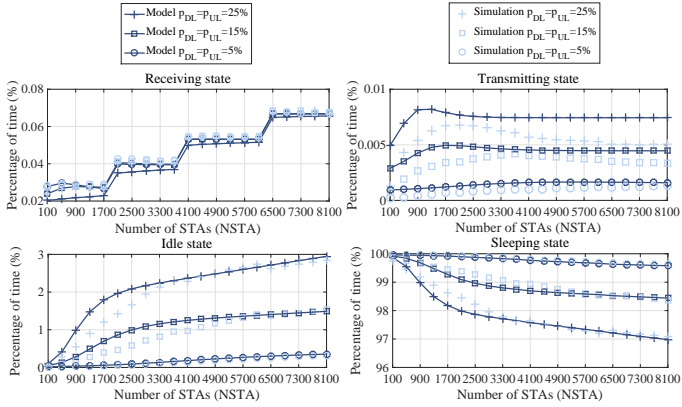}
\caption{Percentage of consumption in each state for different packet generation rates in an indoor scenario \textit{(solid line: $p_{\text{DL}}=0.25, p_{\text{UL}}=0.25$, dashed line: $p_{\text{DL}}=0.15, p_{\text{UL}}=0.15$, dashdotted line: $p_{\text{DL}}=0.05, p_{\text{UL}}=0.05$)}}
\label{fig:indoor}
\end{figure}

\begin{figure}
\centering
\includegraphics[width=9cm]{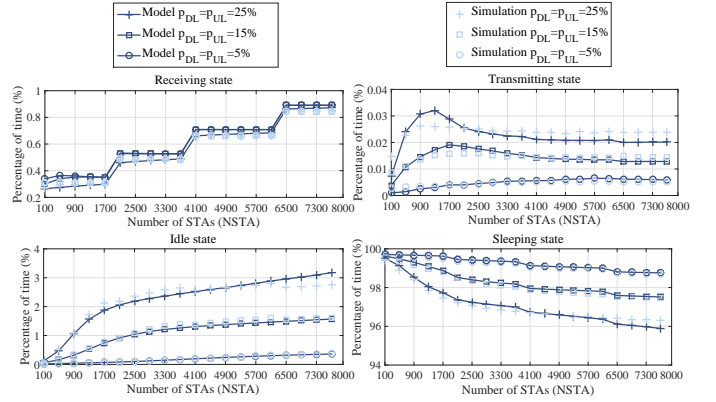}
\caption{Percentage of consumption in each state for different packet generation rates in an outdoor scenario \textit{(solid line: $p_{\text{DL}}=0.25, p_{\text{UL}}=0.25$, dashed line: $p_{\text{DL}}=0.15, p_{\text{UL}}=0.15$, dashdotted line: $p_{\text{DL}}=0.05, p_{\text{UL}}=0.05$)}}
\label{fig:outdoor}
\end{figure}

First simulation results show the percentage of time that an IEEE 802.11ah transceiver remains in each of its possible states (receiving, transmitting, idle, and sleeping) for three different traffic patterns: $p_{\text{DL}}=p_{\text{UL}}=\lbrace 25\%, 15\%, 5 \% \rbrace$; and two scenarios: a $100 m \ x \ 100 m$ indoor network (Figure \ref{fig:indoor}) and a $1000 m \ x \ 1000 m$ outdoor network (Figure \ref{fig:outdoor}), based on the models presented in subsection \ref{scenario}.

One can observe that analytical results in both scenarios are similar to those obtained by simulation. Moreover, it is worth noting that the highest similarity is achieved for low traffic loads. It should be taken into account that, unlike the model, the simulator has a buffer that allows to each STA to retransmit those packets not properly transmitted at their corresponding TIM period. When the traffic load increases, the network behaviour slightly differs from the simulator. The higher the generated traffic, the higher the number of STAs competing and, consequently, the higher the number of collisions. 

When the traffic load increases, differences between the model results and the simulations also do. This difference is mainly caused by the increase of the collision probability, affected by the higher number of STAs competing for the channel. As the simulator has a buffer storing all those non-transmitted packets, the number of active STAs competing for the channel at the next DTIM period will be higher. The higher the collisions, the lower the energy consumed at the transmitting state. For that reason the simulator reduces its transmitting consumption, compared to the results achieved by the model. 
 
Furthermore, the results show that, on average, a STA remains more than  95\% of time in the sleeping state. This fact shows that the IEEE 802.11ah power saving mechanisms allow to reduce the energy consumption by giving the opportunity to the STAs of remaining in a low power state the majority of the operating time.


\subsection{Results from four different application scenarios}

In order to validate the model accuracy in different representative IoT scenarios, we have considered the four use cases (agricultural monitoring, smart metering, industrial automation and animal monitoring) presented in \cite{adame2014IEEE} and summarized at Table \ref{tab:par_scenarios}.

\begin{table}[t!]
\begin{center}
\resizebox {9cm }{!}{
\begin{tabular}{|c|c|c|c|c|}
\hline
 & \textbf{Agricultural } & \textbf{Smart } & \textbf{Industrial } & \textbf{Animal } \\
 & \textbf{ monitoring} & \textbf{ metering} & \textbf{ automation} & \textbf{ monitoring} \\
 \hline
$\mathbf{N_{STA}}$ & 3500 & 15 & 500 & 250 \\
 \hline
 $\mathbf{E[T_{packet}]_{DL}}$ & 240s & 240s & 240s & 240s\\ \hline
 $\mathbf{p_{DL}}$ & 0.67\% & 0.67\% & 0.67\% & 0.67\%\\ \hline
 $\mathbf{E[T_{packet}]_{UL}}$ & 120s & 50s & 180s & 60s\\ \hline
 $\mathbf{p_{UL}}$ & 1.33\% & 3.2\% & 0.89\% & 2.67\%\\ \hline
 \textbf{Area} & 1000x1000m & 8x10m & 250x250m & 1000x1000m \\ \hline
 \textbf{Propagation model} & outdoor & indoor & indoor & outdoor \\ \hline
 
\end{tabular}}
\vspace{5mm}
\caption{Main simulation parameters for four applications scenarios}
 \label{tab:par_scenarios}
\end{center}
\end{table}

\begin{figure}
\centering
\includegraphics[width=9cm]{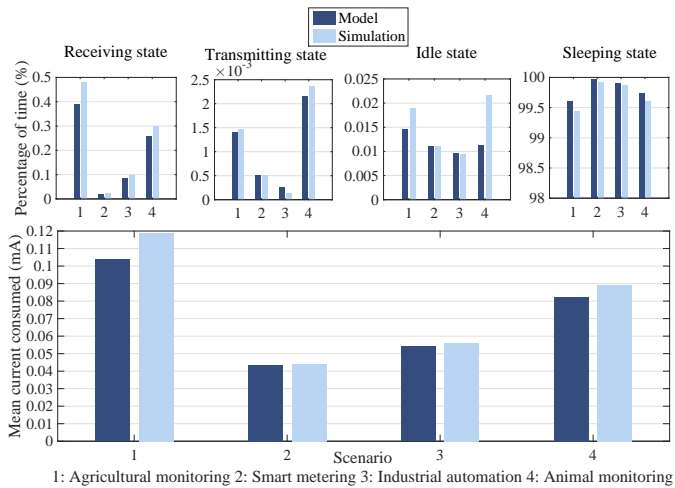}
\caption{Percentage of consumption in each state for the agriculture monitoring scenario.}
\label{fig:comparison}
\end{figure}

The results shown in Figure \ref{fig:comparison} reflect a good accuracy between the model and the simulator in terms of mean current consumed by a STA. If we compare the current consumed, the highest difference (lower than 0.02 mA) appears in the agricultural scenario. As this scenario has the highest number of stations, the number of packets to transmit is also the highest. In that sense, the high collision probability boost the chances of a packet not being transmitted in its corresponding TIM period. Since this situation is not contemplated in our model, the number of packets transmitted in the simulator are higher, which justifies the slight optimism of our model. Moreover, as in the previous results, a STA remains more than 99\% of time at the sleeping state.

By means of using the proposed model, it is possible to predict the energy that a network will consume and even estimate its overall battery lifetime ($B_{\text{LT}}$): 
\begin{equation}
B_{\text{LT}} = \frac{C}{\frac{t_{\text{RX}}I_{\text{RX}}+t_{\text{TX}}I_{\text{TX}}+t_{\text{ID}}I_{\text{ID}}+t_{\text{SL}}I_{\text{SL}}}{T}}
\label{eq:batt_est}
\end{equation}
where C is the capacity of the battery in mAh, $t_{\rho}$ and $I_{\rho}$ are respectively the time spent and the current consumption at each transceiver state, and $\rho=\{\text{RX}, \text{TX}, \text{ID}, \text{SL}\} $ the four different states. 

\begin{figure}
\centering
\includegraphics[width=9cm]{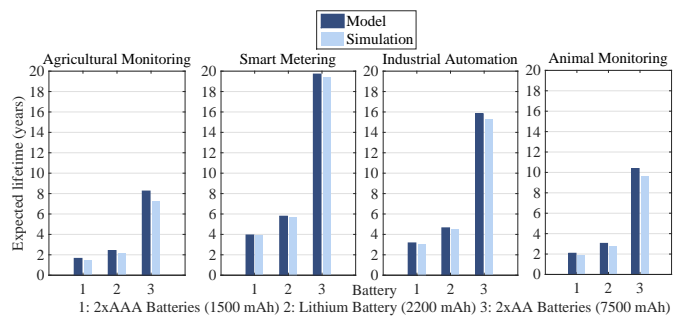}
\caption{Battery lifetime.}
 \label{fig:battery}
\end{figure}

In Figure \ref{fig:battery} we compare the battery lifetime obtained from the simulator and the model. Both values are comparable, although our model is again a little bit more optimistic than the simulator. The major difference obtained is, approximately, 12\% at the agricultural monitoring scenario, which is the one with the largest number of STAs and traffic load.  In order to reduce this possible optimism of our model, one could multiply the results of battery estimation by a factor, e.g. 0.8, in order to reduce this.


\section{Optimization} \label{optimization}

In this section, a performance optimization is done in order to derive the best parameter configuration of an IEEE 802.11ah WLAN for the four representative M2M scenarios introduced before, which cover a wide range of use cases for IEEE 802.11ah WLANs.

The two main parameters that define the channel access in an IEEE 802.11ah network are the number of TIM groups in which stations are distributed ($N_{\text{TIM}}$) and the time between two consecutive DTIM beacons ($T$). Thus, we have evaluated different values of both parameters in order to find the configuration that minimizes the total energy consumption without affecting the probability of successfully transmitting a packet, which must be of $100$ \% in all cases.
 
\subsection{$N_{\text{TIM}}$ optimization}

\begin{figure}
\centering
\includegraphics[width=9cm]{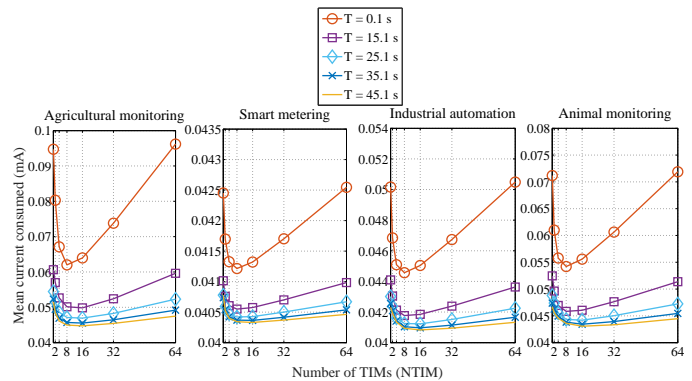}
\caption{Current consumed vs. number of TIM groups ($N_{\text{TIM}}$)}
\label{fig:opt_ntim}
\end{figure}

We can observe in Figure \ref{fig:opt_ntim} that, in terms of current consumed, $N_{\text{TIM}}=8$ is the optimal value for an IEEE 802.11ah STA. Although increasing the number of TIM groups reduces collisions,  the current consumption over the optimal $N_{\text{TIM}}$ value is noticeably affected by the higher size of beacons.

It is worth noting here that the energy consumption is inversely proportional to the time between DTIM beacons ($T$). This is due to the fact that the number of correctly transmitted packets becomes also lower. However, when $T$ is increased, the time between two transmission opportunities also does (a STA can only transmit in its own TIM period), which is a limiting factor in time-critical applications.


\subsection{$T$ optimization}

\begin{figure}
\centering
\includegraphics[width=9cm]{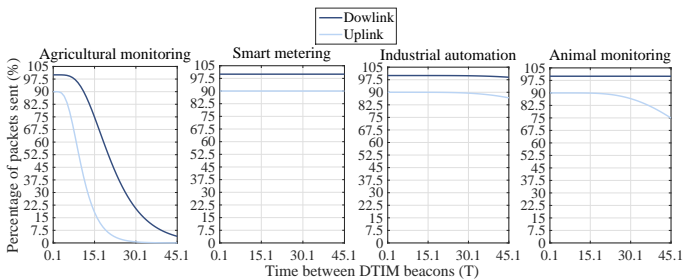}
\caption{Percentage of packets succesfully transmitted vs. time between two consecutive DTIM beacons ($T$ seconds)}
\label{fig:opt_tdtim}
\end{figure}


In Figure \ref{fig:opt_tdtim} we plot the probability of successfully sending a packet when setting $N_{\text{TIM}}=8$ for different values of $T$. This probability diminishes when the time between DTIM periods increases, even if the TIM duration is proportionally extended. As stated in (\ref{Eq:p_psi}), the traffic is generated fixing the time between two consecutive packets. Hence, the higher the value of $T$, the more STAs with pending traffic in each DTIM period. 

To summarize, and taken into account what we have observed in previous results, the $T$ optimal value will be the highest one which ensures the highest probability of successful transmissions. In this occasion, the results reflect that this optimal value highly depends on the scenario analysed: $T=2.4s$ for agricultural monitoring, $T=45.1s$ for smart metering, $T=13.1s$ for industrial automation, and $T=8.1s$ for animal monitoring. One can observe that the $T=1.6s$ value proposed by the standard can be further increased in many scenarios, if we want to reduce the energy consumption without affecting the probability of successful packet transmission.


\subsection{Effects of the optimization}

\begin{figure}
\centering
\includegraphics[width=9cm]{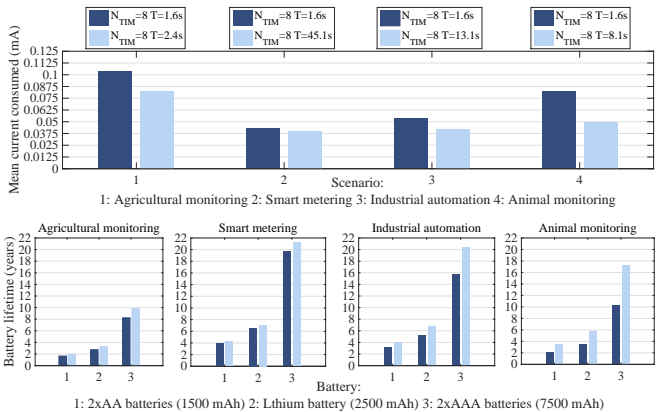}
\caption{Comparison of the model with non-optimized parameters and optimized}
\label{fig:opt_model}
\end{figure}

For each scenario, we compare the mean current consumption and the battery lifetime achieved when using the optimal $T$ and $N_{\text{TIM}}$ values with the predefined ones ($T=1.6s$ and $N_{\text{TIM}}=8$). The results obtained are shown in Figure \ref{fig:opt_model}. The current consumption and the battery lifetime with optimized parameters outperform the performance of the predefined ones. By way of example, current consumption reduction in the animal monitoring scenario is roughly half of the energy consumed with the predefined values. 

Depending on the application scenario, the predefined IEEE 802.11ah values of $T$ and $N_{TIM}$ can be tuned in order to improve the general network performance. Thus, apart from providing a good estimation of the energy consumption in a wide range of scenarios, the model proposed in the current work is an efficient tool to obtain these optimized parameters. 


\section{Conclusions} \label{conclusions}

It is a well-known fact that energy consumption represents one of the most striking challenges in the design and exploitation of WSNs. The study and characterization of this behaviour according to different network conditions such as traffic load or number of stations, as well as other intrinsic network parameters, becomes therefore an essential step previous to further research in energy-saving mechanisms. 

In this work, an analytical model to understand the energy consumption of an IEEE 802.11ah station has been proposed. Its accuracy has been proved by comparing it with simulation results from four representative M2M scenarios. In all of them the model has been an excellent tool to estimate their battery lifetime. 

The effect of varying different network configuration parameters on the whole system has been analysed, showing that the model is able to determine the best values to minimize the current consumption or maximize the success of sending a packet in each scenario. In this regard, increasing the time between DTIM beacons always offers better results in terms of energy consumption. However, it is necessary to take into consideration the trade-off between energy and probability of success when selecting the optimum value. As for the optimization of the number of TIM groups, it has been proven that $N_{\text{TIM}}=8$ minimizes the overall consumed energy. In terms of $T$, its optimized value will highly depend on the scenario. As expected, the higher the traffic generation rate and the number of nodes, the lower the optimal value of $T$.

The current model may be extended in order to include some of the latest IEEE 802.11ah MAC features intended to support energy-efficient communications for sensors. Among them, \cite{park2015ieee} outlines the most notable ones: \textit{Bidirectional TXOP} lets exchange one or more UL and DL packets in a transmission opportunity (TXOP) duration, \textit{NDP CMAC (Null Data Packet Carrying MAC)} reduces overhead of control frames, and \textit{Short MAC Frame} does the same with MAC headers.

Lastly, this model opens the door to further research in the design of advanced sleeping mechanisms which will help to enlarge the battery lifetime of sensor nodes while ensuring proper network operation.

\section*{Acknowledgments}

This work was partially supported by the Spanish and Catalan governments through the projects TEC2012-32354 and SGR-2014-1173, respectively. It has also been funded by the ENTOMATIC FP7-SME-2013 EC project (605073).

\bibliographystyle{IEEEtran}
\bibliography{Bib}

\begin{thebibliography}{10}
\providecommand{\url}[1]{#1}
\csname url@samestyle\endcsname
\providecommand{\newblock}{\relax}
\providecommand{\bibinfo}[2]{#2}
\providecommand{\BIBentrySTDinterwordspacing}{\spaceskip=0pt\relax}
\providecommand{\BIBentryALTinterwordstretchfactor}{4}
\providecommand{\BIBentryALTinterwordspacing}{\spaceskip=\fontdimen2\font plus
\BIBentryALTinterwordstretchfactor\fontdimen3\font minus
  \fontdimen4\font\relax}
\providecommand{\BIBforeignlanguage}[2]{{%
\expandafter\ifx\csname l@#1\endcsname\relax
\typeout{** WARNING: IEEEtran.bst: No hyphenation pattern has been}%
\typeout{** loaded for the language `#1'. Using the pattern for}%
\typeout{** the default language instead.}%
\else
\language=\csname l@#1\endcsname
\fi
#2}}
\providecommand{\BIBdecl}{\relax}
\BIBdecl

\bibitem{machina2014the}
{Machina Research}, ``{The need for low cost, high reach, wide area
  connectivity for the Internet of Things},''
  \url{http://www.neul.com/neul/wp-content/uploads/2014/02/Neul\_Machina\_Research\_White\_Paper.pdf},
  2014, accessed: 2015-11-20.

\bibitem{bellalta2015next}
B.~Bellalta, L.~Bononi, R.~Bruno, and A.~Kassler, ``Next generation ieee 802.11
  wireless local area networks: Current status, future directions and open
  challenges,'' \emph{Computer Communications}, 2015.

\bibitem{aust2012ieee}
S.~Aust, R.~V. Prasad, and I.~G. Niemegeers, ``{IEEE 802.11ah: Advantages in
  Standards and Further Challenges for Sub 1 GHz Wi-Fi},'' in \emph{Proceedings
  of IEEE International Conference on Communications, ICC 2012}, 2012, pp.
  6885--6889.

\bibitem{khorov2015survey}
E.~Khorov, A.~Lyakhov, A.~Krotov, and A.~Guschin, ``{A survey on IEEE 802.11
  ah: An enabling networking technology for smart cities},'' \emph{Computer
  Communications}, vol.~58, pp. 53--69, 2015.

\bibitem{IEEE802.11-12/1084r4}
C.~G. et~al, ``{TIM and Page Segmentation},''
  https://mentor.ieee.org/802.11/dcn/12/11-12-1084-04-00ah-tim-and-page-segmentation.ppt.

\bibitem{IEEE802.11-2012}
``{IEEE Standard for Information technology--Telecommunications and information
  exchange between systems Local and metropolitan area networks--Specific
  requirements Part 11: Wireless LAN Medium Access Control (MAC) and Physical
  Layer (PHY) Specifications},'' \emph{IEEE Std 802.11-2012 (Revision of IEEE
  Std 802.11-2007)}, pp. 1--2793, 2012.

\bibitem{zheng2013performance}
L.~Zheng, L.~Cai, J.~Pan, and M.~Ni, ``{Performance analysis of grouping
  strategy for dense IEEE 802.11 networks},'' in \emph{Global Communications
  Conference (GLOBECOM), 2013 IEEE}, Dec 2013, pp. 219--224.

\bibitem{liu2013power}
R.~P. Liu, G.~Sutton, and I.~Collings, ``{Power save with Offset Listen
  Interval for IEEE 802.11ah Smart Grid communications},'' in
  \emph{Communications (ICC), 2013 IEEE International Conference on}, June
  2013, pp. 4488--4492.

\bibitem{raeesi2014performance}
O.~Raeesi, J.~Pirskanen, A.~Hazmi, T.~Levanen, and M.~Valkama, ``{Performance
  evaluation of IEEE 802.11ah and its restricted access window mechanism},'' in
  \emph{Communications Workshops (ICC), 2014 IEEE International Conference on},
  June 2014, pp. 460--466.

\bibitem{adame2013}
\BIBentryALTinterwordspacing
T.~Adame, A.~Bel, B.~Bellalta, J.~Barceló, J.~Gonzalez, and M.~Oliver,
  ``{Capacity Analysis of IEEE 802.11ah WLANs for M2M Communications.}'' in
  \emph{MACOM}, ser. Lecture Notes in Computer Science, vol. 8310.\hskip 1em
  plus 0.5em minus 0.4em\relax Springer, 2013, pp. 139--155. [Online].
  Available: \url{http://dblp.uni-trier.de/db/conf/macom/macom2013.html}
\BIBentrySTDinterwordspacing

\bibitem{adame2014IEEE}
T.~Adame, A.~Bel, B.~Bellalta, J.~Barcelo, and M.~Oliver, ``{IEEE 802.11AH: the
  WiFi approach for M2M communications},'' \emph{Wireless Communications,
  IEEE}, vol.~21, no.~6, pp. 144--152, December 2014.

\bibitem{EW_subm}
A.~Bel, T.~Adame, B.~Bellalta, J.~Barcelo, J.~Gonzalez, and M.~Oliver,
  ``{CAS}-based channel access protocol for {IEEE} 802.11ah {WLAN}s,'' European
  Wireless 2014; 20th European Wireless Conference; Proceedings of, May 2014.

\bibitem{Park12}
M.~P. et~al, ``Uplink channel access,'' https://
  mentor.ieee.org/802.11/dcn/12/11-12-0606-01-00ah-uplink-channel-access.pptx.

\bibitem{Galcey12}
G.~G. et~al, ``Sectorization for hidden node mitigation,''
  https://mentor.ieee.org/802.11/dcn/12/11-12-0852-00-00ah-sectorization-for-hidden-node-mitigation.pptx.

\bibitem{Hazmi2012}
A.~Hazmi, J.~Rinne, and M.~Valkama, ``{Feasibility Study of IEEE 802.11ah Radio
  Technology for IoT and M2M Use Cases},'' in \emph{Globecom Workshops (GC
  Wkshps), IEEE}, 2012, pp. 1687--1692.

\bibitem{park2015ieee}
M.~Park, ``{IEEE 802.11 ah: sub-1-GHz license-exempt operation for the internet
  of things},'' \emph{{Communications Magazine, IEEE}}, vol.~53, no.~9, pp.
  145--151, 2015.

\end{thebibliography}

\end{document}